\documentclass[11pt]{article}

\usepackage[margin=1.05in]{geometry}
\usepackage{graphicx}
\usepackage{amsmath,amssymb}
\usepackage{booktabs}
\usepackage[colorlinks=true,linkcolor=blue!60!black,citecolor=blue!60!black,urlcolor=blue!60!black]{hyperref}
\usepackage{xcolor}
\usepackage{authblk}

\newcommand{\Teff}{T_{\mathrm{eff}}}
\newcommand{\alphaT}{\alpha_T}
\newcommand{\Rmu}{R_\mu}

\title{Sensitivity of the IceCube atmospheric muon rate to decadal
trends in the effective atmospheric temperature}

\author[1]{Tanvi Krishnan}
\author[2]{Gabriela Barenboim}
\author[1]{Carlos A. Arg\"uelles}
\affil[1]{Harvard University, Department of Physics and Laboratory for Particle Physics and Cosmology, Cambridge, MA 02138, USA}
\affil[2]{Departament de F\'isica Te\'orica and IFIC, Universitat de Val\`encia-CSIC, E-46100, Burjassot, Spain}

\date{\today}

\begin{document}
\maketitle

\begin{abstract}
Decadal changes in the Antarctic stratosphere, generally associated with
ozone recovery and shifts in polar circulation, are of broad geophysical
interest. However, independent long-term measurements of the atmosphere 
above the Antarctic interior are scarce. We assess whether the atmospheric 
muon rate recorded by IceCube could constrain decadal changes in the effective
atmospheric temperature, $\Teff$, a weighted average of temperature through
the atmosphere that is most sensitive to the stratosphere. Using public atmospheric reanalysis data and records of cosmic-ray, solar, and
geomagnetic variability, we reproduce the measured seasonal response of the
IceCube muon rate. We then test whether the same response applies to long-term
temperature changes with a different vertical temperature profile. 
The limiting requirement is detector stability rather than counting statistics:
an unmodeled muon rate drift of $1\%\,\mathrm{decade^{-1}}$ would mimic a temperature
trend of $2.8\,\mathrm{K\,decade^{-1}}$. The long-term rate must therefore be
stable to approximately $0.3\%\,\mathrm{decade^{-1}}$. Within this budget, snow accumulation contributions can be constrained. For a 14-year record, atmospheric 
variability gives a one-sided 95\% CL statistical sensitivity of $1.6 \,\text{K}\,\text{decade}^{-1} $. 
Given this sensitivity, the $\sim0.9\,\text{K}\,\text{decade}^{-1} $ cooling indicated by the 2011--2025 atmospheric 
records would be constrained at the 95\% confidence level after approximately 22 years of measurement.
\end{abstract}

\section{Introduction}
\label{sec:intro}

Polar ozone recovery and associated changes in stratospheric circulation 
produce measurable changes in the Antarctic stratosphere on decadal timescales.
However, the interior of Antarctica is one of the least instrumented regions
of the atmosphere, making independent measurements particularly valuable.
Cosmic-ray muon detectors can provide one such measurement. Although they are 
designed to detect particles rather than measure atmospheric properties, their
muon rates are affected by the temperature of the atmosphere above the detector.
Seasonal variations in the atmospheric muon rate are a well-established consequence
of stratospheric temperature changes. When the stratosphere warms, its density 
decreases at a given altitude and its scale height increases. Charged
pions and kaons produced in cosmic-ray air showers then have a greater 
probability of decaying into muons before undergoing further interactions. 
This makes the muon flux an indirect measure of the temperature in the atmosphere
above the detector~\cite{Barrett1952,Adamson2010,
Grashorn2010,GaisserBook}. 

IceCube is a cubic-kilometer neutrino observatory located at the geographic South
Pole. The detector consists of more than 5000 optical modules embedded in the
Antarctic ice at depths between $1450$ and $2450$\,m~\cite{ICInst}. It was 
designed to detect neutrinos through the Cherenkov light emitted by charged
particles produced in neutrino interactions. In addition to these neutrino events, IceCube
also records a much larger rate of downgoing atmospheric muons.
These muons are produced in cosmic-ray air showers above the detector and then pass
through the Antarctic ice before reaching the detector. Although they are typically
treated as a background in neutrino analyses, their rate varies with the temperature
of the atmosphere above the detector and can therefore be used to monitor
changes in atmospheric temperature. 

The strength of this temperature response can be quantified using the effective temperature
coefficient, $\alpha_T$.  For IceCube, $\alpha_T=0.750\pm0.003$~\cite{Tilav2019}, meaning that a $1\%$ change in the 
effective atmospheric temperature is associated with an approximately $0.75\%$ change in the 
observed muon rate. The detector records events at an average rate of approximately $2.15,\mathrm{kHz}$, 
dominated by muons. Seasonal temperature changes produce an approximately $8\%$ variation in the 
muon rate. This demonstrates that IceCube can monitor atmospheric
temperature changes on seasonal timescales. 
In this paper, we investigate whether the same muon signal can reveal
longer-term changes in the atmospheric temperature over Antarctica. 

Atmospheric reanalyses are long-term reconstructions produced by combining available atmospheric
observations with numerical weather models. They provide one of the most complete available
records of Antarctic temperature, but their reliability is limited in regions where observations are sparse,
including the Antarctic interior. A long-term record from IceCube would not replace these reconstructions. Instead, it would provide an independent measurement of the effective atmospheric temperature, derived from the observed muon rate rather than from the observations and model used to construct the reanalysis. Comparing the two could help determine whether a long-term change is present in both measurements or is specific to the assumptions of a particular method.

We use public atmospheric reanalysis data, neutron-monitor
measurements, and solar and geomagnetic records, together with a calculation of
atmospheric muon production, to test whether the IceCube muon rate could
constrain decadal trends in $\Teff$. We first compare our effective temperature calculation with the
published seasonal response of the muon rate. We then use it to study how the
rate changes under long-term atmospheric temperature variations. We also examine systematic
effects that could mimic or hide a decadal trend, including solar modulation,
atmospheric variability, detector drift, and snow accumulation above the
detector. Finally, we estimate the statistical sensitivity of the method and
the detector stability needed for a long-term measurement. We use simulated
muon rates rather than the measured IceCube rate series, so this study tests the
method and its requirements rather than measuring a temperature trend from
IceCube data.

The rest of this paper is organized as follows. Section~\ref{sec:formalism}
defines the effective atmospheric temperature and describes the data and
simulation inputs. Section~\ref{sec:validation} tests the seasonal temperature
response and compares it with the response to long-term temperature changes.
Section~\ref{sec:solar} estimates solar modulation at the primary energies
relevant for IceCube and describes the solar and geomagnetic records used.
Section~\ref{sec:pseudo} builds simulated muon-rate measurements and tests
whether long-term temperature changes can be recovered from them.
Section~\ref{sec:sensitivity} calculates the statistical sensitivity and the
detector stability needed for a measurement, including the effect of snow
accumulation. Finally, Section~\ref{sec:conclusions} discusses the implications
of these results and the necessary steps to apply the method to the observed 
IceCube muon-rate time series.

\section{Methods and inputs}
\label{sec:formalism}

\subsection{Temperature dependence of the muon rate}

The effect of atmospheric temperature on the muon rate is described by the effective atmospheric
temperature, $\Teff$. This quantity is a weighted average of the atmospheric temperature profile
$T(X)$, where $X$ is the atmospheric column depth and $W(X)$ is the muon production distribution in IceCube~\cite{Grashorn2010,Verpoest2024}:
\begin{equation}
\Teff = \frac{\int\mathrm{d}X\, W(X)\, T(X)}{\int\mathrm{d}X\, W(X)}.\label{eq:teff}
\end{equation}
To first order, a change in the effective atmospheric temperature produces a
proportional change in the muon rate:
\begin{equation}
    \frac{\delta \Rmu}{\langle\Rmu \rangle}=\alphaT\,\frac{\delta\Teff}{\langle\Teff\rangle} ,
    \label{eq:rate_teff}
\end{equation}
where $\Rmu$ is the muon rate, $\delta\Rmu$ and $\delta\Teff$ denote changes
relative to their respective mean values, and $\alphaT$ is the
effective-temperature coefficient defined above.

To calculate the distribution $W(X)$, we use the analytic method for relating atmospheric temperature to underground muon rates developed by 
Grashorn et al.~\cite{Grashorn2010}. This method includes contributions from 
both pions and kaons and describes how temperature changes at different 
atmospheric depths affect the detected muon rate. We use a muon 
spectral index of $\gamma=1.7$, pion and kaon critical energies of $\epsilon_\pi=114\,\mathrm{GeV}$ 
and $\epsilon_K=851\,\mathrm{GeV}$, and interaction column depths of $\Lambda_N=120$, $\Lambda_\pi=160$, and $\Lambda_K=180\,\mathrm{g\,cm^{-2}}$ for primary nucleons, pions, and kaons, respectively~\cite{Grashorn2010}. 

 The distribution $W(X)$ depends on the energy of the muons that reach the detector, which enters the analytic expression through the threshold $E_{\mathrm{th}}$, the minimum surface-muon energy needed to traverse the overburden and trigger the detector. We evaluate it for vertical trajectories and use $500\,\mathrm{GeV}$, close to the approximately $440\,\mathrm{GeV}$ inferred from the vertical overburden (Sec.~\ref{sec:burial}) and consistent with the approximately $400\,\mathrm{GeV}$ reported for IceCube~\cite{Tilav2019}. We vary the threshold from $300$ to $800\,\mathrm{GeV}$ to quantify how the assumed threshold affects the predicted temperature response.

For IceCube, $W(X)$ is strongly concentrated at low pressures: approximately 
$94\%$ of $\int W\,\mathrm{d}X$ comes from pressures below $300\,\mathrm{hPa}$, which is about $8\,\mathrm{km}$ above sea level, 
and the weight peaks near $120\,\mathrm{hPa}$, about $14\,\mathrm{km}$. The polar tropopause lies near $8\,\mathrm{km}$, 
so the muon rate is predominantly sensitive to the stratosphere. Across the threshold 
range considered here, from $300$ to $800\,\mathrm{GeV}$, the effective sampling altitude changes by 
only about $1.5\,\mathrm{km}$. The muon rate therefore provides a weighted temperature average 
over the atmospheric column rather than information about the temperature at any particular altitude.

The measured effective temperature coefficient for IceCube is $\alphaT = 0.750 \pm 0.003$, 
based on the IceCube data collected in 2012 (Fig.~3a of Ref.~\cite{Tilav2019}). The same study 
obtains $\alphaT \approx 0.84$ from its own calculated rate against the same effective 
temperature series (Fig.~3b), the difference reflecting a slightly larger annual modulation 
in the calculated rate than in the measured one. We use the measured value throughout. 
Our calculation does not model the detector's angular acceptance or the zenith dependence 
of the energy threshold, and uses vertical trajectories to study relative changes in the muon
rate produced by changes in atmospheric temperature.

To implement this vertical approximation, we represent the detector with a vertical overburden of $X_V = 1330\,\mathrm{mwe}$. We obtain the corresponding surface-muon energy 
threshold by treating energy loss through the ice as continuous. 
 The optical modules, trigger, event selection, and event 
reconstruction are not simulated. The overburden is represented by a single value rather 
than the $1450$--$2450\,\mathrm{m}$ range of the instrumented volume, and stochastic
fluctuations in muon energy loss are neglected. 

These simplifications are adequate for calculating relative rate changes
and the ratio of the seasonal and long-term responses considered below.  They are not 
intended to predict the absolute event rate or the full detector response.
They would not be sufficient if the result depended on changes in the detector
response over time. The stability requirement in Sec.~\ref{sec:sensitivity}
is therefore stated as a requirement on the detector's monitoring record,
rather than as a quantity derived from this simplified model.
  
\subsection{Atmospheric and solar data}
\label{sec:data}

We use monthly temperature profiles from the NCEP/NCAR Reanalysis~1 at the 
South Pole grid point. The profiles contain temperatures at 
17 pressure levels between $1000$ and $10\,\mathrm{hPa}$ and extend from 1948
to the present~\cite{Kalnay1996}. Reanalysis combines available observations with a 
numerical weather model to produce temperature 
profiles on a regular grid. We use these profiles to calculate $\Teff$ and analyze
temperature changes over time.
Before doing so, we test whether the reanalysis captures the observed variations in
$\Teff$ at the South Pole. We compare the reanalysis-derived $\Teff$ with values
calculated from 217 months of co-located ozonesonde profiles~\cite{NOAAGML}. 
The two estimates have a correlation coefficient of
$r=0.9975$, showing that the reanalysis captures the month-to-month variations
in $\Teff$. The reanalysis is $0.049\,\mathrm{K}$ warmer on average, and the
RMS difference between the two estimates is $1.22\,\mathrm{K}$. These comparisons
support using NCEP/NCAR to represent the long-term temperature variations
in this study.

For the cascade calculation, we use the empirical NRLMSISE-00 atmospheric 
model rather than the NCEP/NCAR data~\cite{Picone2002}. This is because MCEq requires a self-consistent temperature and density profile 
throughout the atmosphere. The NCEP/NCAR data provide temperatures at 17 pressure 
levels, so using them directly in MCEq would require additional interpolation and 
hydrostatic reconstruction of the density profile. We instead use the NRLMSISE-00 profiles 
evaluated for the South Pole, which provide the complete atmospheric state needed by
MCEq without requiring an additional reconstruction step.

In this analysis, we evaluate NRLMSISE-00 for one average atmosphere
in each calendar month. These twelve profiles describe a climatological
seasonal cycle, not the observed differences between individual years.
We run MCEq on them to calculate the seasonal muon-rate response. We use the NCEP/NCAR 
profiles to calculate the monthly $\Teff$ series from 1948 onward, derive the long-term 
temperature changes, and generate simulated muon-rate measurements. These simulated
measurements allow us to test whether a temperature trend can be recovered from
the muon rate. 

We also use the lower-stratospheric satellite temperature record from Remote
Sensing Systems (RSS)~\cite{MearsWentz2009}. This record is used only as an
independent reference for the long-term temperature changes shown in
Sec.~\ref{sec:sensitivity}. It does not enter the calculation of $\Teff$ or
the simulated muon rates.

The atmospheric temperature is not the only factor that can change the muon
rate. The flux of primary cosmic rays entering the atmosphere also varies with
solar activity, producing a rate change that can mimic or obscure a temperature
trend. We therefore use several solar and geomagnetic records to quantify
this effect.

We use daily measurements from the Oulu neutron monitor (1964--2025) and
monthly averages from the South Pole neutron monitor SOPO (2010--2025), both
obtained through NMDB~\cite{NMDB}, to track changes in the primary cosmic-ray
flux. We use the geomagnetic index $A_p$~\cite{Matzka2021} and the $10.7$-cm solar radio 
flux measured at the Dominion Radio Astrophysical Observatory~\cite{Tapping2013}, 
both taken from the combined GFZ daily index file.  The $A_p$ index
characterizes geomagnetic disturbances, while the $10.7$-cm flux
provides a measure of solar activity. The international sunspot number 
is obtained separately from SILSO~\cite{SILSO}. Finally, we use
the heliospheric modulation potential, a model-derived summary of the solar
modulation of the primary cosmic-ray flux, from the monthly reconstruction of
Usoskin et al. for 1936--2025~\cite{Usoskin2005,Usoskin2011}.

The analysis presented here uses simulated muon rates rather than the IceCube muon
time series. We calculate the muon rate from the atmospheric and cosmic-ray inputs
 described above. The only IceCube quantities used are the published effective temperature
 coefficient, $\alphaT=0.750\pm0.003$, and the average trigger rate of approximately 
 $2.15\,\mathrm{kHz}$~\cite{Tilav2019}. The coefficient sets the temperature-response normalization
 in the pseudo-data and provides a benchmark for the simulation, while the trigger rate 
 sets the expected event count and therefore the size 
 of the Poisson fluctuations. We use the NCEP/NCAR
 temperature series to calculate the expected rate and then add Poisson fluctuations to the event
 counts to generate simulated muon-rate measurements. These simulated measurements are used to 
 test whether a temperature trend can be recovered from the muon rate. Applying the same procedure 
 to the IceCube muon time series is left for future work.

\section{Validation of the temperature response}
\label{sec:validation}
\subsection{Validation of the seasonal response}
\begin{figure}[ht]
  \centering
  \includegraphics[width=0.62\linewidth]{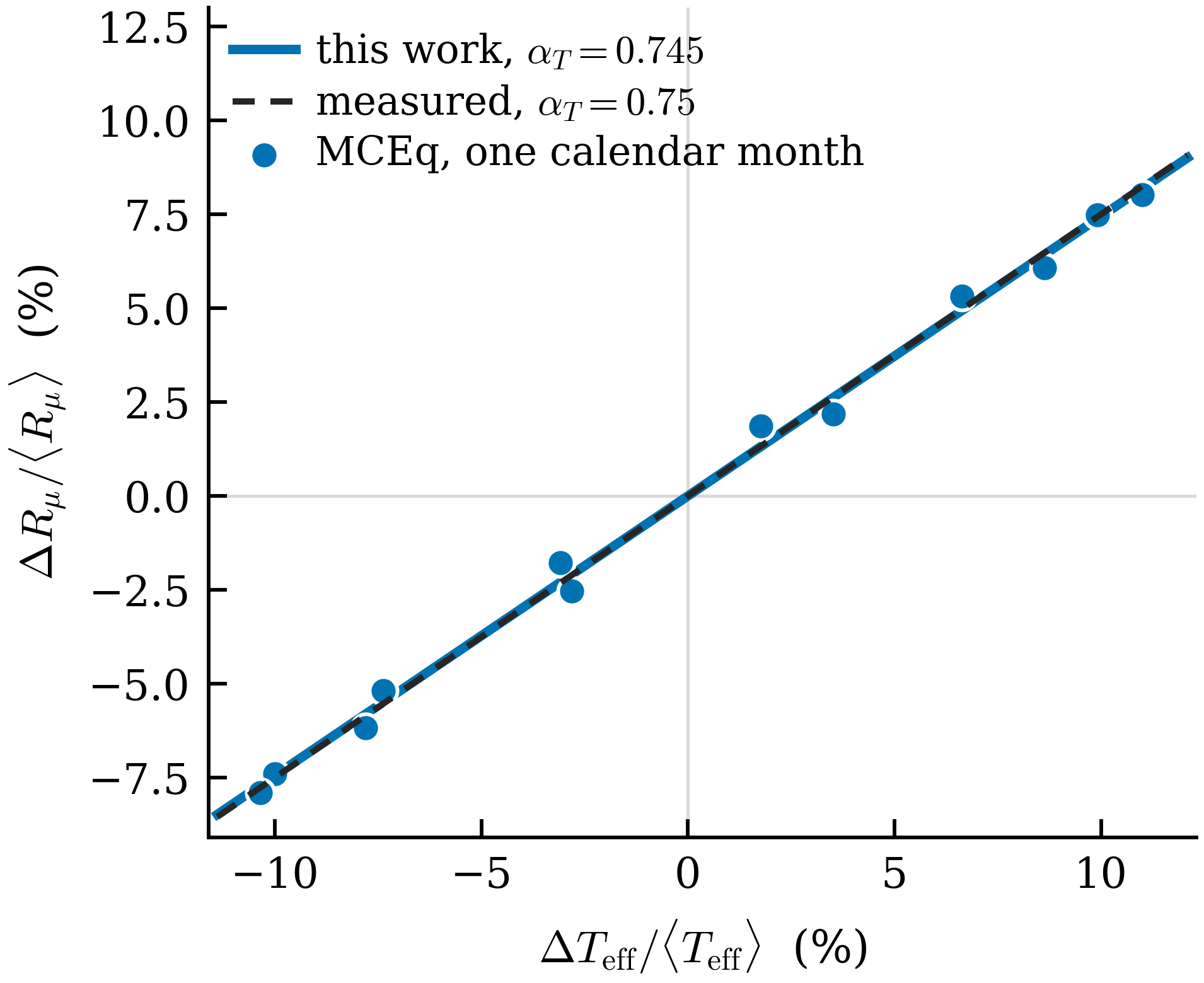}
 \caption{Simulated monthly muon-rate anomalies from the twelve-month mean
plotted against the corresponding anomalies in effective atmospheric
temperature for the twelve representative calendar months. The rates are
calculated with MCEq using SIBYLL-2.3d, H3a, 
and NRLMSISE-00 atmospheric
profiles for the South Pole, with a vertical surface-muon threshold of
$500\,\mathrm{GeV}$. The fitted slope is
$\alphaT=0.745$.}
  \label{fig:alpha}
\end{figure}
We use MCEq~\cite{Fedynitch2015} to calculate the vertical surface-muon
flux for one representative NRLMSISE-00 atmosphere in each calendar month.
The calculation uses SIBYLL-2.3d~\cite{Riehn2020} and the H3a primary
cosmic-ray model~\cite{Gaisser2012}. For each atmosphere, we calculate both
the muon rate and $\Teff$ using the same atmospheric profile and the distribution
$W(X)$. Because NRLMSISE-00 provides the atmosphere through its
density profile, we recover the corresponding temperature profile using
hydrostatic balance. This ensures that the seasonal variations in the muon
rate and $\Teff$ are calculated consistently.

After subtracting the twelve-month mean from the muon rate and from $\Teff$,
we fit the twelve monthly anomalies shown in Fig.~\ref{fig:alpha}. The fit
gives
\begin{equation}
  \alphaT^{\mathrm{MCEq}}(\theta{=}0,\,E_{\mathrm{th}}{=}500\,\mathrm{GeV})
  = 0.745,
  \qquad
  \alphaT^{\mathrm{MCEq}}(300\text{--}800\,\mathrm{GeV})
  = 0.673\text{--}0.805.
\end{equation}
The nominal value is close to the measured coefficient,
$\alphaT=0.750\pm0.003$~\cite{Tilav2019}. The two coefficients are not
identical detector measurements: the published value includes IceCube's full
energy and angular acceptance, whereas our calculation uses vertical muons
and a single energy threshold. 

 After subtracting the twelve-month mean from the muon rate and from $\Teff$, 
 we fit the twelve monthly anomalies shown in Fig.~\ref{fig:alpha}. The fit gives
\begin{equation}
  \alphaT^{\mathrm{MCEq}}(\theta{=}0,\,E_{\mathrm{th}}{=}500\,\mathrm{GeV})
  = 0.745,
  \qquad
  \alphaT^{\mathrm{MCEq}}(300\text{--}800\,\mathrm{GeV})
  = 0.673\text{--}0.805.
\end{equation}
The nominal value is close to the measured coefficient, $\alphaT=0.750\pm0.003$~\cite{Tilav2019}. 
The monthly rate and $\Teff$ anomalies have a correlation coefficient of $r = 0.998$, 
and the simulated peak-to-peak seasonal rate change is $15.9\%$, compared with 
approximately $16\%$ measured by IceCube~\cite{Tilav2019}. Taken together, the 
coupling, the correlation, and the seasonal amplitude show that the calculation 
reproduces the observed seasonal temperature response.
  
  The closeness of the nominal coupling to the measured value should not be read as 
  a precise validation, because the two are not identical quantities. The published value 
  includes IceCube's full energy and angular acceptance, whereas our calculation uses 
  vertical muons and a single energy threshold.  At fixed muon energy, $\alphaT$ decreases 
  toward the horizon, because mesons with inclined trajectories travel farther through thin air,
  giving them a higher probability of decay. On the other hand, the ice overburden 
  raises the effective threshold with zenith angle, which increases $\alphaT$. The two effects 
  partly cancel in the all-sky average. The more robust statement is 
  the threshold range $0.673$--$0.805$, which contains 
  the measured value and whose upper end approaches the $\alphaT \approx 0.84$ obtained 
  from the calculation in Ref.~\cite{Tilav2019}.
 
\subsection{From seasonal variations to decadal temperature trends}
\label{sec:shape}
The seasonal calibration is based on the vertical temperature difference between
the representative January and July atmospheres. The long-term temperature
profile is derived independently from the 2011--2025 NCEP/NCAR series. Because
the muon rate depends on temperature at different atmospheric depths through
the distribution  $W(X)$, the seasonal coupling need not apply exactly to
a temperature change with a different vertical profile. We therefore compare
the seasonal and long-term profiles and test how their different vertical
profiles affect the inferred coupling.

The temperature trend fitted to the NCEP/NCAR series indicates cooling of the
stratosphere by $0.6$--$1.4\,\mathrm{K}\,\mathrm{decade}^{-1}$, with the
largest cooling near $50\,\mathrm{hPa}$, and warming of the troposphere by
$0.3$--$0.8\,\mathrm{K}\,\mathrm{decade}^{-1}$. By contrast, the
January-minus-July seasonal difference is positive at every pressure level.
We compare these temperature-perturbation profiles with a uniform temperature offset. 
Because the NCEP/NCAR reanalysis has no levels at pressures below $10\,\mathrm{hPa}$,
the trend profile must be extrapolated there. This region contributes
$8.3\%$ of $\int W\,\mathrm{d}X$. We consider two extrapolations: holding the
trend fixed at its value at $10\,\mathrm{hPa}$ or tapering it to zero at higher
altitudes. The seasonal profile is taken from the full NRLMSISE-00 atmosphere.

We normalize each temperature-perturbation profile to produce the same change in $\Teff$
and apply it to a full atmospheric profile reconstructed using hydrostatic balance.
We use the same calculation for the perturbed and unperturbed atmospheres,
so common normalization factors cancel in the rate ratios. The reconstructed
unperturbed rate differs from the original NRLMSISE-00 rate by only $0.02\%$.
If the response had a single fixed weighting kernel, all three profiles would
give the same coupling. Their spread shown in Table~\ref{tab:shape} 
therefore measures the dependence of the coupling on the vertical temperature profile.

\begin{table}[ht]
  \centering
  \caption{Effective coupling for three temperature-change profiles, each normalized
  to the same $\Delta\Teff$ and evaluated using January and July baseline atmospheres.
  The range for the NCEP/NCAR trend profile spans the two extrapolations used for pressures below 10\,hPa.}
  \label{tab:shape}
  \begin{tabular}{lcc}
  \toprule
  perturbation shape & $\alpha_{\mathrm{eff}}$ (January)&
  $\alpha_{\mathrm{eff}}$ (July)\\
  \midrule
  seasonal (calibration profile)& 0.724 & 0.745 \\
  NCEP/NCAR trend profile& 0.747--0.756& 0.760--0.772\\
  uniform offset& 0.816& 0.823\\
  \bottomrule
  \end{tabular}
\end{table}
The response is linear to better than $1\%$ for temperature changes of
$1$--$3\,\mathrm{K}$. Changing the baseline atmosphere from January to July
changes the coupling by $1.7\%$. Because all profiles are
normalized to the same change in $\Teff$, the comparison isolates the effect
of their vertical patterns. We therefore compare the response to the
long-term NCEP/NCAR pattern with the response to the January--July seasonal
pattern:
\begin{equation}
  \frac{\alpha_{\mathrm{trend}}}{\alpha_{\mathrm{seasonal}}}
  \;=\; 1.03 \pm 0.02\,\mathrm{(baseline)} \pm 0.01\,\mathrm{(extrapolation)} .
  \label{eq:shape}
\end{equation}
Here, $\alpha_{\mathrm{trend}}$ and $\alpha_{\mathrm{seasonal}}$ denote the
responses to the long-term and seasonal temperature patterns, respectively.
The first uncertainty reflects the difference between the January and July baseline atmospheres.
The second reflects the two extrapolations above $10\,\mathrm{hPa}$.

The long-term temperature pattern therefore produces a muon-rate response
approximately $3\%$ larger than the seasonal pattern. This difference is small
relative to the sensitivity limits considered below, so the measured seasonal
coefficient, $\alphaT=0.750\pm0.003$~\cite{Tilav2019}, is sufficient to set the
response normalization in the pseudo-data. For a future precision measurement,
the corresponding long-term coupling should be taken as approximately
$1.03\,\alphaT$. Using the seasonal coefficient without this adjustment would
overestimate the inferred long-term temperature change by about $3\%$.

\section{Solar effects on the muon rate}
\label{sec:solar}

Solar activity changes the flux of primary cosmic rays entering the atmosphere
and can therefore change the atmospheric muon rate. This variation could mimic
or obscure a long-term temperature signal. Because the analysis covers roughly
one $11$-year solar cycle, the solar contribution cannot be assumed to average
out. We first estimate its size at the primary energies relevant for IceCube.
We then introduce the neutron-monitor and solar-activity records used later
to account for solar variability when inferring long-term temperature changes
from the muon rate.

\subsection{Solar modulation at IceCube energies}

\begin{figure}[t]
  \centering
  \includegraphics[width=0.63\linewidth]{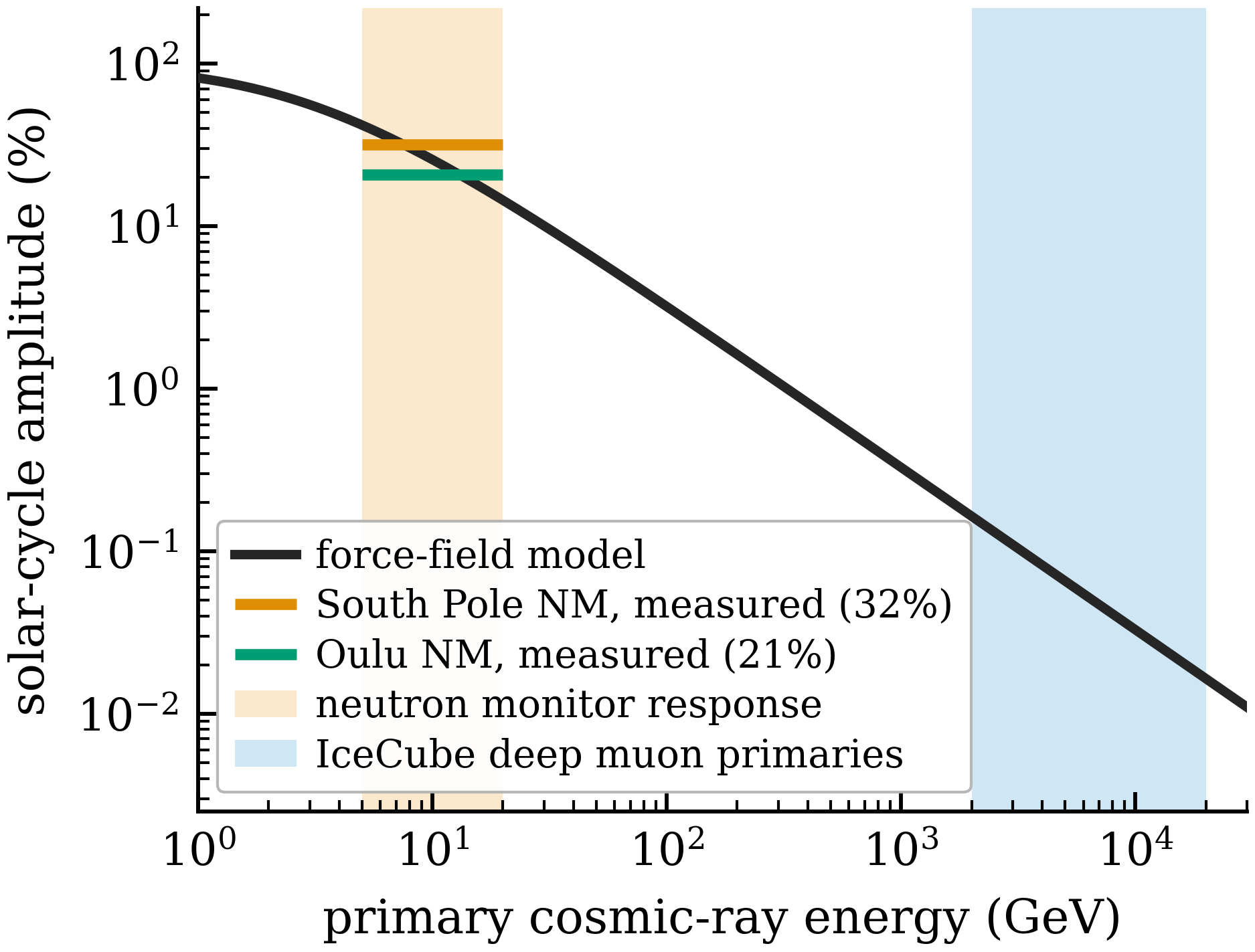}
\caption{
Peak-to-peak fractional modulation of the integrated primary cosmic-ray flux
above a lower energy threshold $E_{\min}$ in the force-field model. The
horizontal bars show the modulation measured by the Oulu and South Pole
neutron monitors over 2010--2025. Their approximate $5$--$20\,\mathrm{GeV}$
response range is shaded in orange. The blue shaded band marks the $2$--$20\,\mathrm{TeV}$ range of minimum primary
energies used to bracket the solar modulation. It does not represent the full
energy distribution of the parent cosmic rays.
}
  \label{fig:solar}
\end{figure}

The IceCube deep-muon sample requires surface-muon energies above approximately
$400\,\mathrm{GeV}$~\cite{Tilav2019}. The parent cosmic rays span a broad
energy distribution rather than a narrow interval. To estimate the effect of
solar modulation at the relevant energies, we use the force-field
model~\cite{GleesonAxford1968}:
\begin{equation}
  J(E) \;=\; J_{\mathrm{LIS}}(E + \phi)\,
  \frac{E\,(E + 2m_p)}
  {(E + \phi)(E + \phi + 2m_p)},
  \label{eq:forcefield}
\end{equation}
where  $E$ is the primary proton kinetic energy, $J(E)$ is the differential
proton intensity at Earth, $J_{\mathrm{LIS}}$ is the
cosmic-ray flux just outside the heliosphere, and $m_p$ is the proton mass.
The heliospheric modulation potential $\phi$ describes the suppression of
cosmic rays by the Sun's magnetic environment. At these energies, the
fractional modulation decreases approximately as $\phi/E$.

Using the reconstructed monthly $\phi(t)$~\cite{Usoskin2005,Usoskin2011}, we
calculate the integrated primary cosmic-ray flux above minimum primary energies
spanning $2$--$20\,\mathrm{TeV}$. For each minimum primary energy, the flux is
integrated up to $10^7\,\mathrm{GeV}$ and evaluated at the minimum and maximum
values of $\phi$ observed during 2011--2025. The resulting peak-to-peak
modulation spans
\begin{equation}
  \left(\frac{\Delta \Phi}{\Phi}\right)_{\!\odot}
  = 0.01\text{--}0.13\%,
\end{equation}
where $\Phi$ denotes the integrated primary cosmic-ray flux. The range reflects
the assumed minimum primary energy and the observed variation in $\phi$, from
$269$ to $1162\,\mathrm{MV}$. This calculation does not include the
energy-dependent muon yield from each primary cosmic ray, so it bounds the size
of the solar effect but does not predict the full modulation
of the muon rate. 

This modulation is roughly two to three orders of magnitude 
smaller than the variations measured by the Oulu and South Pole 
neutron monitors, as shown in Fig.~\ref{fig:solar}, because 
IceCube's deep muons are produced by much higher-energy 
primary cosmic rays. This result establishes the scale of solar 
modulation at the primary energies
relevant for IceCube. Its possible effect on the inferred long-term temperature
change is evaluated later using simulated muon rates.

\subsection{Solar activity records}
\label{sec:solar_records}

 We consider several records as possible measures of solar variability: the
 Oulu and South Pole neutron monitors, both obtained through NMDB~\cite{NMDB}, 
 the 10.7-cm solar radio flux (F10.7) measured at the Dominion Radio 
 Astrophysical Observatory~\cite{Tapping2013}, and the international sunspot 
 number from SILSO~\cite{SILSO}. The South Pole monitor is hereafter denoted SOPO. 
 The neutron monitors measure the arriving cosmic-ray
flux directly, whereas $F_{10.7}$ and the sunspot number provide indirect
measures of solar activity. Oulu provides the longer record needed to compare
the timing of cosmic-ray and solar-activity variations, while SOPO overlaps
with the IceCube data-taking period.
 
The two neutron monitors agree closely over their 191 common months, with
$r=0.996$, and both are anticorrelated with the solar-activity records, with
$|r|\simeq0.85$. To compare the timing of these variations, we shift the
$F_{10.7}$ record relative to Oulu and calculate their correlation. The
correlation has a broad maximum at $143\,\mathrm{d}$ for the full
1964--2025 record and at $239\,\mathrm{d}$ for the IceCube data-taking period, 2011--2025.
Because the preferred lag changes between the two intervals and neither
correlation has a sharply defined maximum, the lag is not fixed by the data
with high precision.

We also examine the geomagnetic index $A_p$ as a separate proxy for
geomagnetic disturbances. Unlike the other records, $A_p$ is neither a
cosmic-ray measurement nor a direct measure of solar activity.

Together, these time series provide alternative ways to represent variations in
the muon rate that are unrelated to atmospheric temperature. The next section
introduces the simulated-rate model used to test whether long-term temperature
changes can be recovered from the muon rate in the presence of these variations.

\section{Recovering long-term temperature changes from simulated muon rates}
\label{sec:pseudo}

Having established the temperature response and estimated the scale of solar
modulation, we now test whether long-term temperature changes can be recovered
from simulated muon-rate measurements.

We construct the simulated rate from the reanalysis-derived effective
temperature series and the measured seasonal coupling:
\begin{equation}
  \frac{R(t)}{\langle R\rangle}
  =
  1
  + \alphaT\,\frac{\delta\Teff(t)}{\langle\Teff\rangle}
  + s_\odot(t)
  + \frac{D}{100}\,t_{\mathrm{dec}}
  + \varepsilon(t).
  \label{eq:pseudo}
\end{equation}
Here, $R(t)$ is the simulated muon rate, $\langle R\rangle$ is its mean
value, and $\delta\Teff(t)=\Teff(t)-\langle\Teff\rangle$ is the effective
temperature anomaly. The temperature series contains 180
monthly values from 2011 to 2025, with
$\langle\Teff\rangle=211.5\,\mathrm{K}$ and a peak-to-peak seasonal variation
of $42.9\,\mathrm{K}$. We use the measured seasonal coupling
$\alphaT=0.750$~\cite{Tilav2019} to set the normalization of the temperature
response. The term $s_\odot(t)$ represents the solar contribution, $D$ is an
injected detector drift in percent per decade, and $\varepsilon(t)$ represents
Poisson fluctuations. The time coordinate $t_{\mathrm{dec}}$ is measured
in decades from the midpoint of the 2011--2025 interval. We first validate the recovery 
of the temperature response and injected drift. We then test whether the choice 
of solar record affects the inferred temperature trend.

\subsection{Constructing and validating simulated rates}

The random term $\varepsilon(t)$ in Eq.~\eqref{eq:pseudo} includes Poisson counting fluctuations
corresponding to the approximately $2.15\,\mathrm{kHz}$ IceCube trigger
rate~\cite{Tilav2019}. For a one-month bin, this corresponds to a 
fractional rate uncertainty of only about $10^{-5}$. We also run the fit with
additional uncorrelated rate fluctuations to test whether the fitted uncertainties
remain properly calibrated when the noise exceeds the Poisson level.

We use the ordinary least squares method to regress the simulated rate on the
effective-temperature anomaly, the solar contribution, and a linear drift.
Because neighboring monthly residuals may be correlated, we use a first-order
autoregressive correction known as AR(1), following Ref.~\cite{Santer2000}.
This correction describes the residual correlation with a single lag-one
coefficient, which we estimate directly from the fit residuals. To report
long-term trends, we first reconstruct the monthly temperatures and then average
them within calendar years before fitting a linear trend.

In the baseline simulation, with no injected drift or additional rate
fluctuations, the fit recovers the injected $\alphaT$ without measurable bias. The trend in
reconstructed temperature agrees with that of the input temperature series
to better than $10^{-3}\,\mathrm{K\,decade^{-1}}$. When a drift is present in the 
simulated rate but omitted from the fit, it produces a spurious temperature trend,

\begin{equation}
  \Delta T_{\mathrm{fake}} \;=\;
  \frac{D}{100}\,\frac{\langle\Teff\rangle}{\alphaT}
  \;\approx\; 2.8\ \mathrm{K\,decade^{-1}\ per\ 1\%\,decade^{-1}}.
  \label{eq:drift}
\end{equation}
Equation~\eqref{eq:drift} gives the expected spurious trend if the entire
unmodeled drift is absorbed by the temperature coefficient. To test this
prediction, we generated pseudo-data with known injected drift values and
fitted each dataset without a drift term while retaining the solar term. The
resulting spurious temperature trends were $94\%$ of the values predicted by
Eq.~\eqref{eq:drift}, because the solar term absorbed a small fraction of the
drift. When the drift term was included in the fit, the injected $D$ was
recovered without bias and the temperature trend was restored.

We generated 300 realizations with additional uncorrelated rate fluctuations of
amplitude $\sigma_{\mathrm{sys}}=5\times10^{-4}$. The fitted values of
$\alphaT$ and $D$ remained unbiased, while their scatter was $12\%$ and $8\%$
larger than the reported uncertainties, respectively. Thus, the additional 
fluctuations do not appreciably bias the recovered
temperature trend, but they make the formal uncertainties on the fitted
parameters slightly too small.

\subsection{Effect of the solar-record choice}

To test whether the choice of solar record affects the inferred temperature
trend, we generate pseudo-data with a solar contribution calculated directly
from the force-field-modulated primary cosmic-ray flux in
Eq.~\eqref{eq:forcefield}. The modulation is driven by the reconstructed
$\phi(t)$ rather than by any solar record. We then refit the same pseudo-data
separately using SOPO, Oulu, $F_{10.7}$ at zero lag, $F_{10.7}$ shifted by the
fitted IceCube-era lag, the sunspot number, $A_p$, and no solar record.

Comparing these fits, including the fit with no solar record, changes the
recovered temperature trend by at most
$8\times10^{-4}\,\mathrm{K\,decade^{-1}}$. The corresponding changes in the
drift coefficient and $\alphaT$ are $0.022\%\,\mathrm{decade^{-1}}$ and
$2\times10^{-4}$, respectively. Linear scaling indicates that the choice of
solar record would affect the recovered temperature trend at the
$0.05\,\mathrm{K\,decade^{-1}}$ level only if the solar modulation amplitude
were approximately $8\%$, or sixty times larger than the estimated physical
amplitude. We therefore find that the choice of solar record has a negligible
effect at IceCube energies, although we retain one in the fits because the data
span only about one solar cycle. At lower thresholds, where the parent cosmic
rays have energies in the GeV rather than TeV range, solar modulation would be
much larger.

\section{Statistical sensitivity and detector stability}
\label{sec:sensitivity}

At the primary energies relevant for IceCube, solar modulation effects are small 
compared with the other sources of variation considered here. Atmospheric variability
determines how precisely a finite record can constrain a temperature trend,
whereas detector drift can mimic such a trend and must be controlled
independently. We first quantify the statistical sensitivity and then determine
the detector stability required for a meaningful long-term measurement.

\subsection{Statistical sensitivity}
\begin{figure*}[t]
  \centering
  \includegraphics[width=\linewidth]{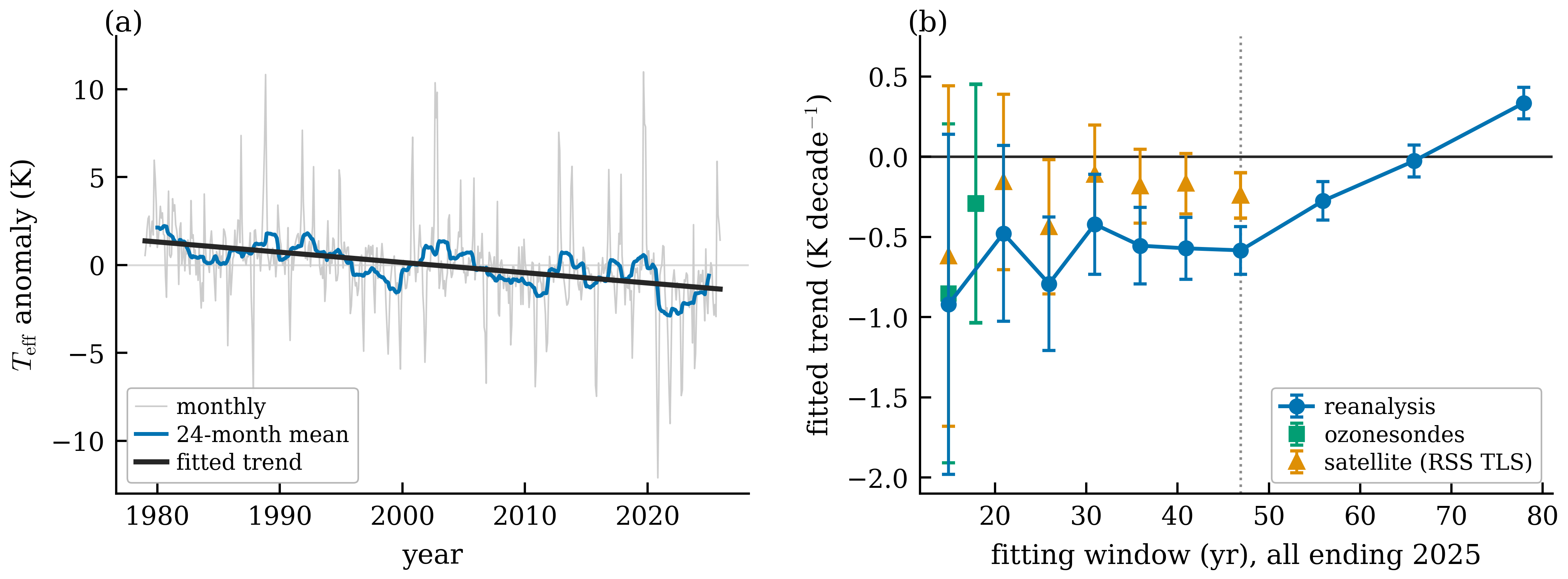}
  \caption{(a)~Monthly anomaly in reanalysis $\Teff$ at the South Pole from
1979--2025, together with its 24-month running mean. The fitted trend is
$-0.58 \pm 0.15$\,K\,decade$^{-1}$, with its uncertainty adjusted for
correlation between neighboring monthly residuals. (b)~Linear trend estimates
as a function of window length for the reanalysis, ozonesonde, and RSS
lower-stratospheric satellite records~\cite{MearsWentz2009}. All windows end in
2025. The dotted line marks the 47-year window shown in panel (a). The
geophysical trend depends on the window length and can change sign, so a
muon-derived trend must be quoted for a specified window.}
  \label{fig:trend}
\end{figure*}
The polar stratosphere exhibits substantial month-to-month 
 variability, as shown in Fig.~\ref{fig:trend}a. Individual monthly anomalies have a 
 standard deviation of $2.5$\,K, whereas the fitted long-term trend is $-0.58 \pm 0.15$\,K\, decade$^{-1}$ 
 over 1979--2025. The trend is therefore small compared with the individual 
monthly variations and can be constrained only with a long record. The precision 
depends on both the variance of the temperature series and its month-to-month 
autocorrelation. Propagating these quantities through the trend fit gives the 
sensitivities in Table~\ref{tab:floors}, quoted as one-sided 95\%~CL sensitivities for a 
null trend. These should be compared with the trend over IceCube's own observing 
period rather than with the full reanalysis record: from 2011 to 2025 the reanalysis 
indicates approximately $0.9\,\mathrm{K\,decade^{-1}}$ of cooling, with an uncertainty 
comparable to the trend itself, and the ozonesonde and satellite records show the 
same sign.

\begin{table}[t]
  \centering
  \caption{Statistical limits on a decadal $\Teff$ trend, as one-sided
  95\%~CL upper limits.}
  \label{tab:floors}
  \begin{tabular}{lcccc}
  \toprule
  baseline & 14\,yr & 20\,yr & 30\,yr & 40\,yr \\
  \midrule
  95\%~CL sensitivity (K\,decade$^{-1}$) & 1.6 & 0.93 & 0.51 & 0.33 \\
  \bottomrule
  \end{tabular}
\end{table}

The values in Table~\ref{tab:floors} are statistical sensitivities. They include
atmospheric variance and Poisson counting noise, which contributes only about
$7\times10^{-4}$ of the atmospheric variance. They do not include IceCube's
own non-Poisson rate variability, which cannot be estimated without the
IceCube muon-rate time series.

Any temperature trend inferred from these data must be reported for a specified
time window. The South Pole temperature trend depends strongly on the window
length and can change sign, as shown in Fig.~\ref{fig:trend}b, as it is 
strongly affected by ozone-related variability rather than
representing a monotonic greenhouse-driven trend. The trend should
therefore be reported for a stated window or as an upper limit,
rather than as a prediction of future temperature change. The muon measurement
also has no vertical resolution, as noted in Sec.~\ref{sec:formalism}, and
therefore represents a weighted average over the atmospheric column.

The AR(1) correction assumes that correlations between residuals are short-lived
and can be described by a single lag-one coefficient. We tested for longer-range
dependence using three estimators of the Hurst exponent $H$. Because these
estimators are biased high for records of this length, we compared the
satellite-era value with a simulated null distribution. The 95th percentile of
that distribution is $H=0.79$, while the satellite-era record gives $H=0.73$.
We therefore find no evidence that long-range dependence requires replacing the
AR(1) treatment used to calculate the statistical sensitivities.

This test has limited power at the available record length: against 
genuine long-range dependence at $H=0.75$, it rejects in only 
about one-third of simulations. The measured value should 
therefore be interpreted as showing no detectable long-range 
dependence, rather than as ruling it out or proving that the noise 
follows an AR(1) model. At that level of dependence, the 40-year 
statistical sensitivity would increase from $0.33$ to 
$0.43\,\mathrm{K\,decade^{-1}}$, which remains below the approximately 
$0.9\,\mathrm{K\,decade^{-1}}$ cooling over the IceCube data-taking period.

\subsection{Required detector stability}

Equation~\eqref{eq:drift} gives a linear conversion between an unmodeled
rate drift and an apparent temperature trend. From 2011 to 2025, the
reanalysis indicates cooling of approximately
$0.9\,\mathrm{K\,decade^{-1}}$, with the ozonesonde and satellite records
showing the same sign. Keeping an unmodeled detector drift below a comparable apparent 
trend requires
\begin{equation}
  |D| \;\lesssim\; 0.3\%\ \mathrm{decade}^{-1}.
  \label{eq:requirement}
\end{equation}
The 14-year statistical sensitivity is $1.6\,\mathrm{K\,decade^{-1}}$, which exceeds the approximately
$0.9\,\mathrm{K\,decade^{-1}}$ trend indicated by the atmospheric records. Atmospheric variability alone
would therefore prevent a 95\% constraint on this trend over that baseline. The statistical
sensitivity falls below this trend at a baseline of roughly 22 years.

A fixed detector drift produces a trend-like signal that cannot be distinguished
from a genuine long-term temperature change using the rate series alone. Longer
records reduce statistical uncertainty but do not remove this systematic
ambiguity. Detector stability therefore sets the practical limit once the
statistical sensitivity becomes smaller than the atmospheric trend of interest.

The closest published quantities to the stability measure needed here are the
track detection probability and the gain calibration. The track detection
probability is the per-module probability that a muon track passing within
$100\,\mathrm{m}$ produces a detected hit. For the standard modules that
dominate the muon trigger, its mean changed by $-0.1\%$ between 2012 and 2015
~\cite{ICInst}. If this rate of change continued for a decade, it would
correspond through Eq.~\eqref{eq:drift} to approximately
$0.9\,\mathrm{K\,decade^{-1}}$ of apparent cooling. Two calibration epochs
cannot establish whether that change was sustained. The gain calibrations also
varied from year to year at the $0.1$--$0.5\%$ level, and a cumulative
$-0.3\%$ shift over five years was identified and corrected~\cite{ICInst}.
These measurements show that changes of the relevant size can be monitored,
but neither quantity is the trigger rate itself. Gain measures charge
calibration, while track detection probability measures per-module efficiency
rather than the array-wide counting rate. Establishing the stability needed
for this measurement therefore requires the trigger-rate record to be
analyzed directly as a time series.

\subsection{Snow accumulation}
\label{sec:burial}

\begin{figure}[ht]
  \centering
  \includegraphics[width=0.6\linewidth]{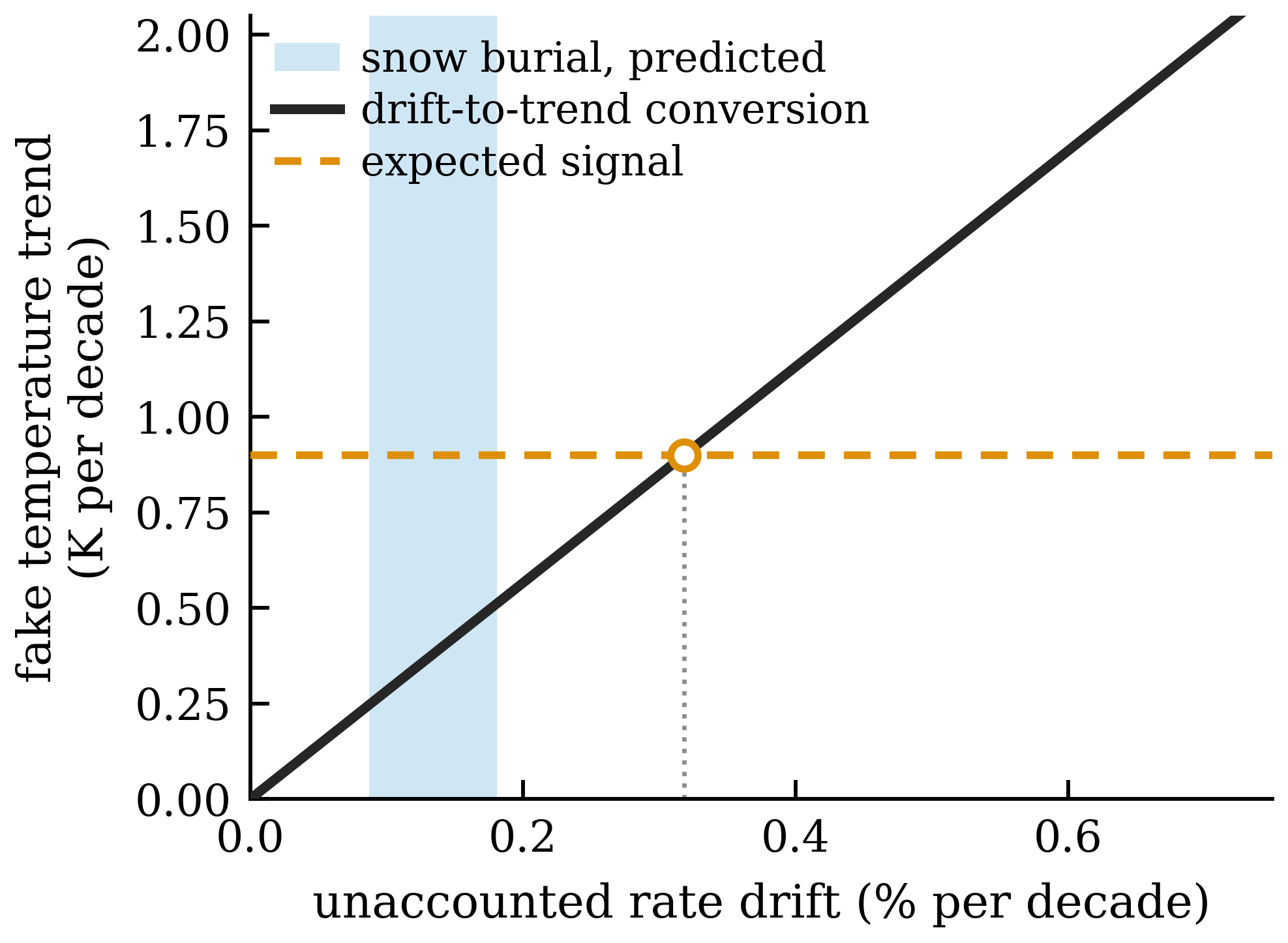}
  \caption{Conversion between an unmodeled detector-rate drift and the
  apparent temperature trend from Eq.~\eqref{eq:drift}. The figure compares
  the observed IceCube-era temperature trend with the predicted contribution from
  snow accumulation and shows the corresponding detector-stability requirement.}
  \label{fig:equiv}
\end{figure}
Snow accumulation produces a muon-rate change that can be estimated in
advance. Snow accumulates at the South Pole at $8$\,cm water equivalent per
year~\cite{Lilien2018,MosleyThompson1999}. The overburden above the in-ice
detector therefore increases by approximately $0.8$\,m\,w.e.\ per decade,
from a representative vertical column of approximately $1330$\,m\,w.e.
Using the modeled average muon energy loss in ice, this overburden corresponds
to a surface-muon threshold of approximately $440\,\mathrm{GeV}$. Without
accounting for ice flow, the additional snow would increase this threshold by
up to $0.33\,\mathrm{GeV\,decade^{-1}}$. Near this threshold, the simulated
muon rate has a local logarithmic slope
$-\mathrm{d}\ln R_\mu(>E_{\mathrm{th}})/\mathrm{d}\ln E_{\mathrm{th}}=2.40$.

Horizontal ice flow reduces the fraction of surface accumulation that adds to
the overburden above the detector. In the simple ice-flow model, this fraction
is $1-X_V/H$, where $X_V$ is the detector depth and $H$ is the full ice-column
depth. For the representative column used here, the fraction is approximately
$0.5$. We therefore bracket the effect of ice flow between a factor of $0.5$
and the limiting case of no reduction, with a factor of $1$. The resulting
snow-induced rate change is
\begin{equation}
  \left(\frac{\dot R}{R}\right)_{\rm snow}
  = -(0.087\text{--}0.181)\ \%\,\mathrm{decade}^{-1}.
  \label{eq:burial}
\end{equation}
The range is dominated by the uncertainty in the ice-flow reduction rather than by
the muon energy-loss calculation. Through
Eq.~\eqref{eq:drift}, the snow-induced rate change corresponds to
$0.24$--$0.51\,\mathrm{K\,decade^{-1}}$ of apparent cooling. It has the same
sign as the atmospheric trend and represents roughly one quarter to
one half of its magnitude. It must therefore be included in any trend analysis, as shown in Fig.~\ref{fig:equiv}.

The snow-induced rate change is a substantial part of the
detector-stability requirement in Eq.~\eqref{eq:requirement}. Because its
magnitude is predicted rather than fitted, it should enter the analysis as a
constrained contribution rather than as a free parameter. Allowing for its
full predicted range leaves roughly
$0.21\%\,\mathrm{decade^{-1}}$ for other unmodeled changes in the measured
detector rate. It also provides a diagnostic of detector stability. If the
measured rate includes the predicted snow-induced change, a stable detector
should show a fitted long-term rate change consistent with that prediction.
Any additional long-term rate change should be consistent with zero if the
detector is otherwise stable.

\section{Discussion and conclusions}
\label{sec:conclusions}

These results show that the IceCube atmospheric muon rate can, in principle,
measure or usefully bound a decadal trend in effective atmospheric temperature
over the South Pole, provided the detector-stability requirement is met. Our
calculation reproduces the measured seasonal response of the muon rate to
atmospheric temperature fluctuations. Applying this response to decadal changes
requires a $3\%$ correction (Sec.~\ref{sec:validation}). Solar modulation
changes the rate by less than one percent at the relevant energies
(Sec.~\ref{sec:solar}), while atmospheric variability gives a one-sided 95\% CL
statistical sensitivity of $1.6\,\mathrm{K\,decade^{-1}}$ for a 14-year record
and $0.33\,\mathrm{K\,decade^{-1}}$ for a 40-year record
(Sec.~\ref{sec:sensitivity}). The statistical sensitivity decreases with
record length, but long-term detector-rate drift does not average down with
additional livetime. An undetected detector-rate drift of $1\%$ per decade
would mimic a temperature trend of $2.8\,\mathrm{K\,decade^{-1}}$, requiring
the detector rate to remain stable at approximately the
$0.3\%\,\mathrm{decade^{-1}}$ level. The predicted snow-induced rate change
spans approximately $0.087$--$0.181\%\,\mathrm{decade^{-1}}$, leaving roughly
$0.21\%\,\mathrm{decade^{-1}}$ for other unmodeled detector drift.

These results also identify the requirements for applying the method to the
measured IceCube rate. The rate should be modeled jointly with the effective
temperature, a solar-activity record, and a linear detector-rate drift.
The solar and drift coefficients should be reported together because they can be
partly degenerate over a single solar cycle. The predicted rate change from
snow accumulation should enter as a constrained contribution rather than as a
free drift parameter. Before being applied to the measured rate, the regression 
should be validated on pseudo-data with
known temperature, solar, and detector-rate contributions.

The method is not specific to IceCube. The relation between muon rates and
effective atmospheric temperature in Eq.~\eqref{eq:teff} applies to other deep
detectors with known overburden, and seasonal temperature responses have been
measured in underground-muon experiments~\cite{Adamson2010,Grashorn2010,Verpoest2024}.  
Deep water detectors such as KM3NeT/ARCA and the planned P-ONE have 
comparable overburdens and could support the same analysis, although their different 
depth, surface elevation and latitude would change the weighting and the resulting coupling.
IceCube is particularly well suited to this application because it has a 
high event rate, a deep overburden that suppresses solar modulation, and a 
site where the muon rate is strongly weighted toward the stratosphere. 
At shallower sites, solar modulation becomes
more important and must be modeled using independent solar-activity records.

A natural refinement would be to replace the simplified detector model with a
calculation that includes the detector's angular acceptance, energy-dependent
threshold, and stochastic muon energy loss. These refinements 
would improve the precision of a future measurement, but are
not required for the present analysis, which uses the published IceCube
seasonal coefficient $\alphaT=0.750$~\cite{Tilav2019} to normalize the
temperature response. More broadly, this work suggests that a telescope 
built to detect neutrinos may also measure the atmosphere above it through the 
muons that reach the detector.

\section*{Acknowledgements}

GB is supported by PID2023-151418NB-I00 funded by MCIU/AEI/10.13039/501100011033/, and by the European ITN project HIDDeN (H2020-MSCA-ITN-2019/860881-HIDDeN).
TK is supported by the U.S. Department of Energy, Office of Science, Office of Advanced Scientific Computing Research, Department of Energy Computational Science Graduate Fellowship under Award Number DE-SC0025528. C.A.A. are supported by the Faculty of Arts and Sciences of Harvard University, the National Science Foundation (NSF, CAREER Grant \#2239795), the John Templeton Foundation (Grant \#63651), the Research Corporation for Science Advancement, and the David \& Lucile Packard Foundation.


\begin{thebibliography}{99}

\bibitem{Barrett1952}
P.~H. Barrett, L.~M. Bollinger, G.~Cocconi, Y.~Eisenberg, and K.~Greisen,
\emph{Interpretation of cosmic-ray measurements far underground},
Rev. Mod. Phys. \textbf{24} (1952) 133.

\bibitem{Adamson2010}
P.~Adamson \emph{et al.} (MINOS Collaboration),
\emph{Observation of muon intensity variations by season with the MINOS far
detector}, Phys. Rev. D \textbf{81} (2010) 012001.

\bibitem{Grashorn2010}
E.~W. Grashorn \emph{et al.},
\emph{The atmospheric charged kaon/pion ratio using seasonal variation methods},
Astropart. Phys. \textbf{33} (2010) 140, arXiv:0909.5382.

\bibitem{GaisserBook}
T.~K. Gaisser, R.~Engel, and E.~Resconi,
\emph{Cosmic Rays and Particle Physics}, 2nd ed., Cambridge University
Press (2016).

\bibitem{ICInst}
M.~G. Aartsen \emph{et al.} (IceCube Collaboration),
\emph{The IceCube Neutrino Observatory: instrumentation and online systems},
JINST \textbf{12} (2017) P03012, arXiv:1612.05093.

\bibitem{Tilav2019}
S.~Tilav, T.~K. Gaisser, D.~Soldin, and P.~Desiati (IceCube Collaboration),
\emph{Seasonal variation of atmospheric muons in IceCube},
PoS(ICRC2019)~894 (2021), arXiv:1909.01406.

\bibitem{Verpoest2024}
S.~Verpoest, D.~Soldin, and P.~Desiati,
\emph{Atmospheric muons and their variations with temperature},
Astropart. Phys. \textbf{161} (2024) 102985, arXiv:2405.08926.

\bibitem{Kalnay1996}
E.~Kalnay \emph{et al.},
\emph{The NCEP/NCAR 40-year reanalysis project},
Bull. Am. Meteorol. Soc. \textbf{77} (1996) 437.

\bibitem{NOAAGML}
NOAA Global Monitoring Laboratory, \emph{South Pole ozonesonde vertical
profiles}, Earth System Research Laboratories, Boulder, CO,
\url{https://gml.noaa.gov/ozwv/ozsondes/}.


\bibitem{Picone2002}
J.~M. Picone, A.~E. Hedin, D.~P. Drob, and A.~C. Aikin,
\emph{NRLMSISE-00 empirical model of the atmosphere: statistical comparisons
and scientific issues},
J. Geophys. Res. \textbf{107} (2002) 1468.

\bibitem{MearsWentz2009}
C.~A. Mears and F.~J. Wentz,
\emph{Construction of the Remote Sensing Systems V3.2 atmospheric
temperature records from the MSU and AMSU microwave sounders},
J. Atmos. Oceanic Technol. \textbf{26} (2009) 1040.

\bibitem{NMDB}
Neutron Monitor Database, \url{https://www.nmdb.eu} (EU FP7 no.~213007).
Oulu station: Sodankyl\"a Geophysical Observatory; South Pole station:
University of Wisconsin--River Falls.

\bibitem{Matzka2021}
J.~Matzka \emph{et al.},
\emph{The geomagnetic Kp index and derived indices of geomagnetic activity},
Space Weather \textbf{19} (2021) e2020SW002641.

\bibitem{Tapping2013}
K.~F. Tapping,
\emph{The 10.7\,cm solar radio flux (F10.7)},
Space Weather \textbf{11} (2013) 394--406,
doi:10.1002/swe.20064.

\bibitem{SILSO}
WDC--SILSO, Royal Observatory of Belgium, Brussels,
\url{https://www.sidc.be/SILSO}.

\bibitem{Usoskin2005}
I.~G. Usoskin, K.~Alanko-Huotari, G.~A. Kovaltsov, and K.~Mursula,
\emph{Heliospheric modulation of cosmic rays: monthly reconstruction for
1951--2004}, J. Geophys. Res. \textbf{110} (2005) A12108.

\bibitem{Usoskin2011}
I.~G. Usoskin, G.~A. Bazilevskaya, and G.~A. Kovaltsov,
\emph{Solar modulation parameter for cosmic rays since 1936 reconstructed
from ground-based neutron monitors and ionization chambers},
J. Geophys. Res. \textbf{116} (2011) A02104. Updated series at
\url{https://cosmicrays.oulu.fi/phi/}.

\bibitem{Fedynitch2015}
A.~Fedynitch, R.~Engel, T.~K. Gaisser, F.~Riehn, and T.~Stanev,
\emph{Calculation of conventional and prompt lepton fluxes at very high
energy}, EPJ Web Conf. \textbf{99} (2015) 08001, arXiv:1503.00544.

\bibitem{Riehn2020}
F.~Riehn, R.~Engel, A.~Fedynitch, T.~K. Gaisser, and T.~Stanev,
\emph{Hadronic interaction model SIBYLL~2.3d and extensive air showers},
Phys. Rev. D \textbf{102} (2020) 063002.

\bibitem{Gaisser2012}
T.~K. Gaisser,
\emph{Spectrum of cosmic-ray nucleons, kaon production, and the atmospheric
muon charge ratio}, Astropart. Phys. \textbf{35} (2012) 801.

\bibitem{GleesonAxford1968}
L.~J. Gleeson and W.~I. Axford,
\emph{Solar modulation of galactic cosmic rays},
Astrophys. J. \textbf{154} (1968) 1011.

\bibitem{Santer2000}
B.~D. Santer \emph{et al.},
\emph{Statistical significance of trends and trend differences in
layer-average atmospheric temperature time series},
J. Geophys. Res. \textbf{105} (2000) 7337.

\bibitem{Lilien2018}
D.~A. Lilien \emph{et al.},
\emph{Holocene ice-flow speedup in the vicinity of the South Pole},
Geophys. Res. Lett. \textbf{45} (2018) 6557.

\bibitem{MosleyThompson1999}
E.~Mosley-Thompson, J.~F. Paskievitch, A.~J. Gow, and L.~G. Thompson,
\emph{Late 20th century increase in South Pole snow accumulation},
J. Geophys. Res. \textbf{104} (1999) 3877.

\end{thebibliography}
\end{document}